# Current-driven domain wall motion along ferromagnetic strips with periodically-modulated perpendicular anisotropy

Luis Sánchez-Tejerina[1], Óscar Alejos[1], Víctor Raposo[2],

and Eduardo Martínez[2,*]

[1]Dpto. Electricidad y Electrónica, University of Valladolid, 47011 Valladolid, Spain.
[2]Dpto. Física Aplicada, University of Salamanca, 37008 Salamanca, Spain.

* corresponding author: edumartinez@usal.es

**Abstract**

The dynamics of magnetic domain walls along ferromagnetic strips with spatially modulated perpendicular magnetic anisotropy is theoretically studied by means of micromagnetic simulations. Ferromagnetic layers with a periodic sawtooth profile of the anisotropy depict a well-defined set of energy minima where the walls are pinned in the absence of external stimuli, and favor the unidirectional propagation of domain walls. The performance of the current-driven domain wall motion along these ratchet-like systems is compared to the field-driven case. Our study indicates that the current-driven domain wall motion exhibits significant improvements with respect to the field-driven case in terms of bit shifting speed and storage density, and therefore, it is suggested for the development of novel devices. The feasibility of these current-driven ratchet devices is studied by means of realistic micromagnetic simulations and supported by a one-dimensional model updated to take into account the periodic sawthooth anisotropy profile. Finally, the current-driven domain wall motion is also evaluated in systems with a triangular modulation of the anisotropy designed to promote the bidirectional shifting of series of walls, a functionality that cannot be achieved by magnetic fields.

**I. INTRODUCTION**

The ability to propagate series of magnetic domain walls (DWs) along ferromagnetic strips is the basis of several proposed logic and memory devices[1,2]. There are some requirements for successful DW-based memory devices, where the information is coded within the magnetic domains between DWs. Firstly, it is needed to nucleate (write operation) a magnetic domain and its DWs[3]. The nucleated or written domain (which acts as the data bit) should be efficiently shifted along the ferromagnetic strip to the location of the reading head. Therefore, other issue is to control the exact pinning positions where the DWs stop after propagation. Recent studies focused on multilayers with inversion structural asymmetry, where a ultrathin ferromagnetic layer (FM) is sandwiched between a heavy metal (HM) and an oxide (Ox)[4,5,6,7,8] have opened new promising ways of efficient DW motion. These systems exhibit a high perpendicular magnetic anisotropy (PMA), and the interfacial Dzyaloshinskii-Moriya interaction (DMI)[9] at the HM/FM interface makes DWs adopt a Néel configuration. These homochiral DWs are efficiently driven by the torque exerted by an injected electrical current through the HM, due to the spin Hall effect (SHE)[6,7]. In contrast, an applied out-of-plane magnetic field drives adjacent DWs in exactly opposite directions; this is the main reason why the manipulation of DWs with electrical currents is so useful.

As mentioned, apart from the high efficient current-driven DW motion, the design of DW-based devices requires to develop an adequate pinning strategy to precisely control the DW positions along the device. In HM/FM/Ox multilayers, adjacent homochiral DWs experience magnetostatic interaction[10,11,12], which imposes a limit in the stored bit density. Moreover, under current pulses, these homochiral DWs can also depict inertia[13] which results in some after-effect DW motion once the current pulse is switched off. The introduction of notches along the strip was initially suggested to fix the DW positions[14]. However, these shape-induced effects were initially designed for systems with in-plane magnetization, while in systems with high PMA this effect is rather weak, and typically induces deformations of the DW[15]. For these reasons, some alternatives have been proposed, such as the control of the DW pinning (and DW nucleation) by means of a tailored PMA. These alternatives include the application of a voltage in an epitaxial magnetic tunnel junction[16], or a strain-mediated coupling in piezoelectric/magnetostrictive bilayer structures[17]. The irradiation of the sample with heavy ions is another procedure[18,19] which can produce an anisotropy landscape along the ferromagnetic strip.

The PMA is known to be reduced by irradiation with highly energetic ions, and therefore, the anisotropy can be controlled very locally at a scale of a few nanometers. Using different doses of local irradiation with heavy ions, Franken *et al.*[20] generated a sawtooth profile, where the local PMA anisotropy ($K_u = K_u(x)$) varies linearly from a minimum ($K_u^-$) to a maximum ($K_u^+$) over a distance $d$, and this tooth is repeated periodically along the ferromagnetic strip. This idea led to a new proposal for a magnetic memory, known as *ratchet memory*, which was studied under the field-driven regime. The periodic sawtooth PMA profile was intended to fix the DW positions and to establish one direction of bit shifting by avoiding backwards DW movement due to the applied field. As adjacent up-down (↑↓) and down-up (↓↑) DWs move in opposite directions under a fixed out-of-plane field orientation, an alternate applied magnetic field (bipolar field pulse) is needed to promote the shifting of the DWs. Moreover, as one DW is driven by the field whereas the other one remains pinned for a fixed field polarity (either along the $+z$ or the $-z$ direction), two teeth are needed to store a single bit of information, a fact which imposes the minimum bit size $b_s = 2d$. In the present work we go deeper into this conception and study theoretically for the first time the DW dynamics in a ratchet memory device by means of unipolar current pulses. Using micromagnetic simulations ($\mu M$), we evaluate the dynamics of DWs under perfect and realistic conditions to explore the feasibility of the proposed current-driven ratchet device. Additionally, the one-dimensional model (1DM) is also updated here to include the effective field accounting for the periodic anisotropy landscape. This 1DM is used to clarify some aspects of the DW dynamics and will be of help in the further development of the proposed devices. We will show that the current-driven mechanism ensures the proper bit shifting along the FM strip. Indeed, it constitutes a much more interesting alternative from the technological point of view to the field-driven basis, since the current promotes the dynamics of all DWs in the same direction. Moreover, the current-driven ratchet also contributes to reduce bit sizes (increasing bit densities) and to speed up the bit shifting. Besides, we also suggest other systems where the PMA is periodically modulated to achieve the bidirectional DW motion, and theoretically analyze the current-driven DW back and forth shifting along them.

The present manuscript is structured as follows: Sec. II presents the system under study along the details of the Micromagnetic model ($\mu M$) and the one-dimensional model (1DM) developed to describe the DW dynamics in systems with periodically modulated PMA. Sec. III.A reviews the main features of the field-driven ratchet and unveils its handicaps when dealing with bit densities and bit shifting times as compared to the current-driven case (Sec. III.B). The current-driven DW shifting along a ratchet strip studied in Sec. IV. Full $\mu M$ simulations and 1DM results are presented for perfect strips, both at zero and at finite temperature. Besides, the effect of disorder is evaluated to provide a more realistic description. Sec. V sketches how to extend the exposed ideas to implement a bi-directional device. Finally, the main conclusions are discussed in Sec. VI.

## II. GEOMETRY AND MODELS

A ratchet FM strip with high PMA sandwiched between a HM and an Ox, as schematically depicted in Fig. 1, is considered. The left graph in Fig. 1 depicts four periods of the sawtooth profile of anisotropy ($K_u = K_u(x)$) along the longitudinal axis of the FM strip ($x$-axis). Within each tooth of the sawtooth the PMA constant increases linearly from $K_u^-$ to $K_u^+$ over a distance $d$, which defines the periodicity of the sawtooth anisotropy profile along the $x$-axis. Each region of length $d$ can pin one DW, which in the absence of external stimuli locates approximately at any of the points where the anisotropy energy is a minimum The word "*approximately*" accounts for the finite width of the DW and the abrupt drop of the anisotropy from the maximum ($K_u^+$) to the minimum ($K_u^+$) for the ratchet case.

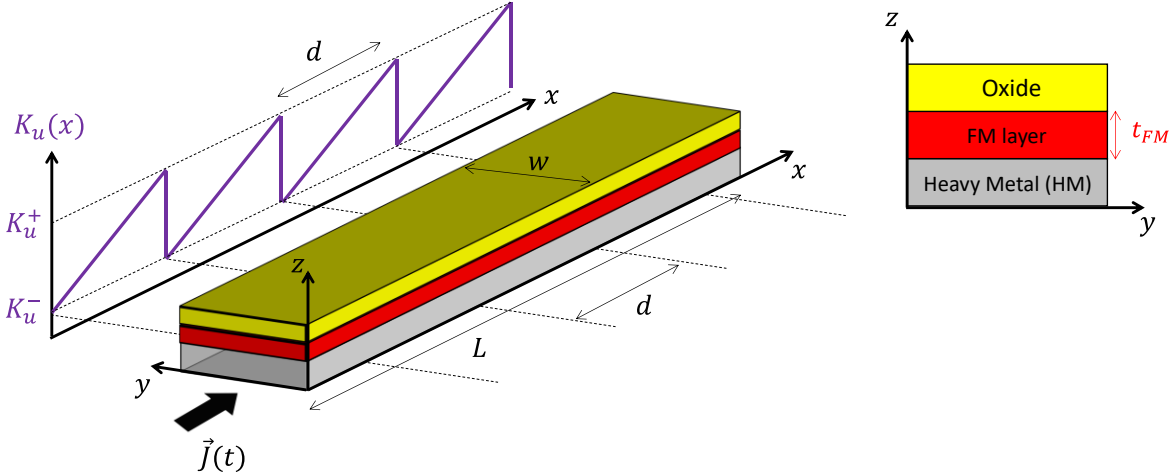

**FIG. 1.** Description of the system under study. A stack consisting of a FM strip of width $w$ and thickness $t_{FM}$ sandwiched between a HM and an Oxide is considered. The FM layer exhibits high PMA, but such an anisotropy is tailored to present a sawtooth profile along the longitudinal direction ($x$). This ratchet profile is defined by two extreme values $K_u^+$ and $K_u^-$, and its periodicity is given by a distance $d$. All minima define subsequent equilibrium positions where DWs get pinned.

### II.A Micromagnetic model ($\mu M$)

The temporal evolution of the normalized magnetization $\vec{m}(\vec{r},t) = \vec{M}(\vec{r},t)/M_s$ is described by the Landau-Lifshitz-Gilbert (LLG) equation augmented by the spin-orbit torques[5,7,21,22] (SOT) and thermal fluctuations[23]:

$$\frac{d\vec{m}}{dt} = -\gamma_0 \vec{m} \times (\vec{H}_{eff} + \vec{H}_{th}) + \alpha \vec{m} \times \frac{d\vec{m}}{dt} + \vec{\tau}_{SO} \quad (1)$$

where $\gamma_0$ and $\alpha$ are the gyromagnetic ratio and the damping parameter respectively. $\vec{H}_{eff}$ is the effective field, which includes the exchange, the perpendicular magnetic anisotropy, the magnetostatic and the Zeeman interactions, along with the DMI[9]. For the field-driven case, the applied field is pointing along the out-of-plane direction, $\vec{B}_{ext} = B_e(t)\vec{u}_z$ with $B_e(t) = \mu_0 H_e(t)$. $\vec{\tau}_{SO}$ represents the spin-orbit torque (SOT), which in general includes two contributions: $\vec{\tau}_{SO} = \vec{\tau}_{SL} + \vec{\tau}_{FL}$. The first one, $\vec{\tau}_{SL}$, is the Slonczewski-like (or damping-like) SOT term, which is given by $\vec{\tau}_{SL} = -\gamma_0 \vec{m} \times \vec{H}_{SL}$, where $\vec{H}_{SL} = H_{SL}^0 (\vec{\sigma} \times \vec{m})$ is the Slonczewski-like effective field[7,24]. Here, $\vec{\sigma} = \vec{u}_z \times \vec{u}_J$ is the unit vector along the direction of the polarization of the spin current generated by the spin Hall effect (SHE)[25,26] in the HM, which is orthogonal to both the direction of the electric current $\vec{u}_J = \vec{u}_x$ and the vector $\vec{u}_z$ standing for the normal to the HM/FM interface. $H_{SL}^0 = \hbar \theta_{SH} J_a(t) / (2\mu_0 e M_s t_{FM})$ determines the strength of the SHE[24], where $\hbar$ is the Planck constant, $e < 0$ is the electron charge, $\mu_0$ is the vacuum permeability, $\theta_{SH}$ is the spin Hall angle, and $J_a(t)$ is the magnitude of the current density $\vec{J}(t) = J_a(t)\vec{u}_x$ injected through the HM. The second contribution to the SOT ($\vec{\tau}_{FL}$) is the Field-like SOT, which is expressed as $\vec{\tau}_{FL} = -\gamma_0 \vec{m} \times \vec{H}_{FL}$, with $\vec{H}_{FL}$ being the corresponding effective field given by $\vec{H}_{FL} = H_{FL}^0 \vec{\sigma}$. The magnitude of the Field-like SOT effective field is $H_{FL}^0 = kH_{SL}^0$, where the factor $k$ parameterizes $H_{FL}^0$ in terms of the Slonczewski-like effective field $H_{SL}^0$. $\vec{H}_{th}$ is the thermal field included as a Gaussian-distributed random field[23,27].

The study of the motion of a DW in such a system requires solving this equation by means of full micromagnetic simulations ($\mu M$). In order to carry out those simulations, typical material parameters of HM/FM/Ox multilayers commonly found in the literature[9,21,28,29] have been considered: saturation magnetization $M_s = 1.1$ MA/m, exchange constant $A_{ex} = 16$ pJ/m and a Gilbert damping parameter $\alpha = 0.5$. The values $K_u^+ = 1.27$ MJ/m$^3$, $K_u^- = 1.0$ MJ/m$^3$ and $d = 128$ nm define the anisotropy landscape. The anisotropy variation has been chosen to be compatible with the reported in recent experimental works[18]. Except otherwise indicated, a uniform DMI parameter $D = 1$ mJ/m$^2$ has been considered[30]. This value is sufficiently high to induce the formation of homochiral Néel DWs and, therefore, to allow efficient current-driven DW movement by means of the SHE[7]. The considered spin Hall angle is $\theta_{SH} = 0.1$. The adopted magnetic parameters mimic a multilayer of Pt(3)/Co(0.6)/AlO(2) where the corresponding thicknesses between brackets are in given in nm[9,21,28,29]. Although it is now well established that the main driving force on the DW in theses systems comes from the Slonczewski-like SOT term, some additional simulations were carried out considering also a finite Field-like SOT contribution, which was equal in magnitude to the Slonczewski-like SOT ($|k| = 1$). It was verified that this Field-like SOT ($|k| = 1$) does not significantly modify the results obtained in the absence of Field-Like SOT ($k = 0$), and therefore, the presented results were computed assuming $k = 0$. Indeed, the main driving force on the DW is the Slonczewski-like SOT. A FM strip with a cross section $w \times t_{FM} = 128$ nm $\times$ 0.6 nm has been analyzed. $\mu M$ simulations have been performed with MuMax3 package[31] using a time 1 ps. The in-plane side of the computational cells is $\Delta x = \Delta y = 1$ nm. Except the contrary is said, the presented results were obtained at zero temperature. Simulations at room temperature were performed with a fixed time step $\Delta t = 0.1$ ps. Several tests were performed with reduced cell sizes and time steps to assess the numerical validity of the presented results.

Part of the simulations was carried out by considering perfect samples, without imperfections nor defects. Additionally, other parts were computed under realistic conditions (see Sec. IV.C). In order to take into account the effects of disorder due to imperfections and defects in a realistic way, we assume that the easy axis anisotropy direction ($\vec{u}_K = \vec{u}_K(\vec{r}_G)$) is distributed among a length scale defined by a grain size[13]. The average size of the grains is 10 nm. Despite the fact that the direction of the uniaxial anisotropy of each grain is mainly directed along the perpendicular direction (z-axis), a small in-plane component lower than 5% is randomly generated over the grains. The inter-grain exchange stiffness parameters was assumed to be equal to the exchange parameter $A_{ex}$. It was verified that a 25% reduction of the inter-grain exchange constant does not modify the presented results. Besides, a random edge roughness pattern is considered along both edges of the FM strip with a characteristic size of 4 nm. Although other ways to account for imperfections could be adopted[32], we selected this one based on previous studies, which properly describe other experimental observations[13,12].

### II.B One dimensional model (1DM)

The one dimensional model (1DM) assumes that the DW profile can be described by the Bloch's ansatz[33], $\theta(x,t) = 2\tan^{-1}\left[\exp\left(Q\frac{x-q(t)}{\Delta}\right)\right]$, where $\theta$ is the angle of magnetization with respect to the out-of-plane direction, z-axis, $\Delta$ is the DW width and $Q = \pm 1$ correspond to the up-down (↑↓, $Q = +1$,) and down-up (↓↑, $Q = -1$) DW configurations. The DW dynamics can be described by means of the DW position ($q$) along the strip axis (x-axis), and the internal DW angle ($\Phi$), which is defined with respect to the $+x$-axis. The 1DM has been developed by several authors to account for and describe the field-driven and current-driven DW dynamics in different systems[9,28,10]. Here we have developed it to analyze the DW dynamics in systems depicting a modulated profile of the perpendicular magnetic anisotropy (PMA) along the strip axis, $K_u = K_u(x)$. In general, the resulting 1DM equations from Eq. (1) can be expressed as

$$\alpha \frac{\dot{q}}{\Delta} + Q\dot{\Phi} = -\frac{\gamma_0}{2\mu_0 M_s} \frac{\partial \sigma}{\partial q} + \gamma_0 Q H_{th} + \gamma_0 Q \frac{\pi}{2} H_{SH} \cos\Phi \qquad (2)$$

$$Q\frac{\dot{q}}{\Delta} - \alpha\dot{\Phi} = \frac{\gamma_0}{2\mu_0 M_s \Delta} \frac{\partial \sigma}{\partial \varphi} - \gamma_0 \frac{\pi}{2} H_{FL} \cos\Phi \qquad (3)$$

where the top dot notation represents the time derivative ($\dot{q} \equiv \frac{dq}{dt}$). The term $H_{SH} \equiv H_{SL}^0 = \frac{\hbar \theta_{SH} J_a(t)}{2\mu_0 e M_s t_{FM}}$ is the Slonczewskii-like term associated to the SHE, and $H_{FL} = kH_{SL}^0$ is the Field-like effective field. Thermal fluctuations are accounted by the thermal term $H_{th}(t)$. $\sigma$ is the total areal energy density[9] which results from the integration of the volume energy density $\omega$ over the strip axis, $\sigma = \int_{-\infty}^{+\infty} \omega\, dx$, and it includes the same interactions as the effective field $\vec{H}_{eff}$ in Eq. (1): exchange, magnetostatic, DMI, PMA and interaction with an external field.

$$\sigma = \frac{2A_{ex}}{\Delta} + \Delta\mu_0 M_s^2 (N_x \cos^2\Phi + N_y \sin^2\Phi - N_z) \qquad (4)$$
$$+ Q\pi D \cos\Phi - 2Q\mu_0 M_s H_z q + \sigma_{PMA}$$

where $N_x, N_y, N_z$ are the magnetostatic factors, and $\sigma_{PMA} = \int_{-\infty}^{+\infty} \omega_{PMA} dx$ is the areal energy density due to the PMA, with $\omega_{PMA} = K_u(x) \sin^2\theta(x)$. As here we are interested in strips with modulated PMA along their axis, $K_u = K_u(x)$, this term deserves a particular treatment. Let us initially consider a general periodic PMA profile where $K_u$ increases linearly from $K_u^-$ to $K_u^+$ over a distance $d_1$, and after that it decreases linearly from $K_u^+$ to $K_u^-$ over a distance $d_2$, with $d = d_1 + d_2$ being the period of the PMA profile. This can be expressed as

$$K_u(x) = \begin{cases} K_u^- + \frac{K_u^+ - K_u^-}{d_1}\left(x - \left\lfloor\frac{x}{d}\right\rfloor d\right), & 0 \leq \left(x - \left\lfloor\frac{x}{d}\right\rfloor d\right) < d_1 \\ K_u^+ - \frac{K_u^+ - K_u^-}{d_2}\left(x - \left\lfloor\frac{x}{d}\right\rfloor d - d_1\right), & d_1 \leq \left(x - \left\lfloor\frac{x}{d}\right\rfloor d\right) \leq d \end{cases} \qquad (5)$$

where $\left[\frac{x}{d}\right]$ represents the integer part of $\frac{x}{d}$. Note that here we are interested in strips where the DW width is significantly smaller than the PMA period ($\Delta \ll d$). Under this circumstance and after some algebra, the PMA areal density $\sigma_{PMA} = \int_{-\infty}^{+\infty} \omega_{PMA} dx$ can be expressed as

$$\begin{aligned}\sigma_{PMA} &= +2\Delta K_u^- \\ &+ (K_u^+ - K_u^-) \sum_{n:-1}^{+1} \left[\frac{\Delta^2}{d_1}\log\left(\frac{\cosh\frac{d}{\Delta}a_n}{\cosh\frac{d}{\Delta}\left(a_n - \frac{d_1}{d}\right)}\right) + \frac{\Delta^2}{d_2}\log\left(\frac{\cosh\frac{d}{\Delta}(a_n - 1)}{\cosh\frac{d}{\Delta}\left(a_n - \frac{d_1}{d}\right)}\right)\right]\end{aligned} \quad (6)$$

where $a_n = \left\{\frac{q}{d}\right\} + n$, with the braces $\left\{\frac{q}{d}\right\}$ indicating the fractional part of $\frac{q}{d}$, and $n: -1, 0, +1$. Therefore,

$$\frac{1}{2\mu_0 M_s}\frac{\partial \sigma_{PMA}}{\partial q} \equiv H_{PMA}(q) = H_{PMA}^0 r(q) \quad (7)$$

where $H_{PMA}(q)$ represents the effective field due to the periodic PMA profile, with $H_{PMA}^0 = \frac{(K_u^+ - K_u^-)}{2\mu_0 M_s}$, and

$$r(q) = \sum_{n:-1}^{1}\left[\frac{\Delta}{d_1}\left(\tanh\frac{d}{\Delta}a_n - \tanh\frac{d}{\Delta}\left(a_n - \frac{d_1}{d}\right)\right) + \frac{\Delta}{d_2}\left(\tanh\frac{d}{\Delta}(a_n - 1) - \tanh\frac{d}{\Delta}\left(a_n - \frac{d_1}{d}\right)\right)\right] \quad (8)$$

The general PMA profile $K_u = K_u(x)$ (Eq. (5)), and the corresponding PMA $r(q)$ function (Eq. (7)-(8)) are represented in Fig. 2(a) for $d_1 = 2 d_2$ and $K_u^- = 1$ MJ/m$^3$ and $K_u^+ = 1.27$ MJ/m$^3$. This effective field $H_{PMA}(q) = H_{PMA}^0 r(q)$ drives the DW to the closest accessible energy minima in the absence of driving force: a DW located between $0 \leq q \leq d_1$ is pushed back towards $q \approx 0$ by a negative $H_{PMA}(q)$, whereas a DW between $d_1 \leq q \leq d_1 + d_2$ is forced towards $q \approx d_1 + d_2$ by a positive $H_{PMA}(q)$ (see bottom graph in Fig. 2(a)). The magnitude of this out-of-plane PMA effective field is determined by the slope of the local PMA parameter $K_u(x) = K_u(q)$.

The ratchet case, where $K_u$ periodically increases linearly from $K_u^-$ to $K_u^+$ over a distance $d$ (Fig. 2(b)), can be also obtained from Eq. (5) with $d = d_1$ and $d_2 = 0$. The corresponding PMA $r(q)$ function (Eq. (8)) is shown in the bottom graph of Fig. 2(b). As it can be easily understood, a DW at an intermediate location $0 \leq q \leq d$ is always forced backwards to $q \approx 0$ by the corresponding PMA effective field, $H_{PMA}(q) = H_{PMA}^0 r(q)$.

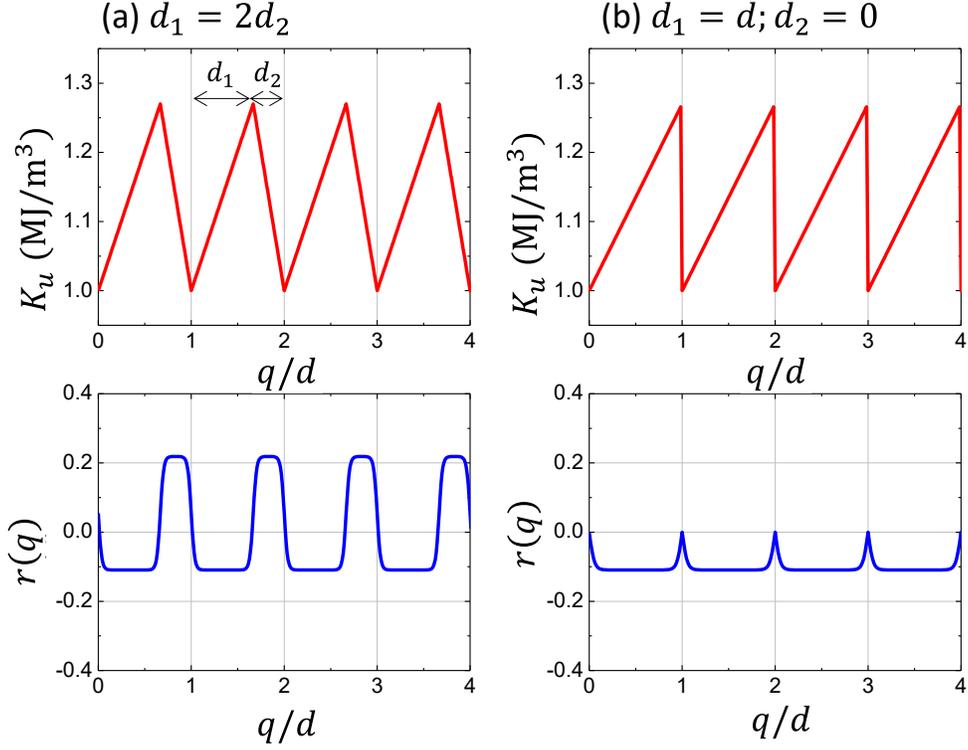

**Figure 2.** Periodic anisotropy profile $K_u = K_u(x)$ and corresponding local pinning PMA field $r(q) \equiv H_{PMA}(q)/H_{PMA}^0$ as function of the DW position $q$. The period of the anisotropy profile is $d = d_1 + d_2$. (a) General case with $d_1 > d_2$. (b) Ratchet case with period $d = d_1$ and $d_2 = 0$. In both cases $K_u^- = 1$ MJ/m³ and $K_u^+ = 1.27$ MJ/m³.

By substituting Eq. (4) and (6) in Eqs. (2) and (3) the 1DM equations can be finally expressed as:

$$\dot{q} = \frac{\gamma_0 \Delta}{1 + \alpha^2} [Q\Omega_A + \alpha \Omega_B] \quad (9)$$

$$\dot{\Phi} = \frac{\gamma_0}{1 + \alpha^2} [-\alpha \Omega_A + Q\Omega_B] \quad (10)$$

where

$$\Omega_A = Q\frac{\pi}{2}H_D \sin\Phi - \frac{H_k}{2}\sin(2\Phi) - \frac{\pi}{2}H_{FL}\cos\Phi \qquad (11)$$

$$\Omega_B = -\frac{\pi}{2}H_{SL}\cos\Phi + QH_z + H_{th}(t) + H_{PMA}(q) \qquad (12)$$

$H_D = \frac{D}{\mu_0 M_s \Delta}$ is the DMI effective field[9,7,24], where $\Delta$ is the DW width which was assumed to be constant ($\Delta = \sqrt{\frac{A}{K_{eff}}}$, where $K_{eff} = \widetilde{K}_u - \frac{1}{2}\mu_0 M_s^2$, with $\widetilde{K}_u = (K_u^+ + K_u^-)/2$). $H_k = M_s(N_x - N_y) \approx M_s N_x$ is the magnetostatic shape anisotropy field, where $N_x = \frac{t_{FM}\log(2)}{\pi\Delta}$ is the magnetostatic factor[34]. Eqs. (9)-(10) are solved numerically using a Runge-Kutta scheme.

## III. RATCHET MEMORY: FIELD-DRIVEN *vs* CURENT-DRIVEN DW SHIFTING.

### III.A. Field-driven case

Before focusing our attention to the current-driven DW along a ratchet device, here we review the main features of the field-driven case, which was initially suggested by Franken *et al*.[20] to develop a magnetic memory device based in the unidirectional motion of trains of DWs along a ratchet strip. A periodic anisotropy sawtooth profile given by Eq. (6) with $d = d_1$ and $d_2 = 0$ (ratchet PMA profile) is considered to study the DW dynamics by means of $\mu M$ simulations. The field-driven case requires bipolar pulses to displace trains of DWs along the same strip because up-down (↑↓) and down-up (↓↑) DWs are driven in opposite directions by the same field pulse. This fact is shown in the example depicted in Fig. 3(b), where three central magnetic domains (↑↓↑: 101, up/down magnetized domains represent 1/0 bits respectively) are separated by down-up (↓↑) and up-down (↑↓) DWs, which are initially located at their equilibrium positions close to the minimum the sawtooth profile ($K_u^-$). Each domain extends over a distance equivalent to the length of two slopes of the PMA profile, the bit size being $b_s = 2d$ in this case. Note that down-up DWs are driven to the right ($x > 0$ direction) by the negative field-pulse ($B_e(t) < 0$ for $0 < t < 2$ ns), whereas up-down DWs remain trapped at their initial locations. Indeed, the strength of the applied field is not sufficiently intense to promote their movement towards the left due to the high energy barrier associated with the sawtooth anisotropy profile. A second positive field pulse ($B_e(t) > 0$ for 2 ns $< t <$ 4 ns) is needed to displace up-down DWs along the positive $x$-axis afterwards (see snapshots for $t \geq 2$ ns in Fig. 3(b)), whereas down-up DWs remain trapped during this second pulse. In this case, where each bit extends over a distance $b_s = 2d$, the bipolar field pulse shown in Fig. 3(b) drives the sequence of bits (↑↓↑: 101) a single tooth ($d$).

In order to illustrate the limitations of the field-driven case in terms of bit density, an analogous study to that shown in Fig. 3(b) is now carried out, but reducing the initial distance between the adjacent DWs to $b_s = d$. The corresponding snapshots shown in Fig. 3(c) indicate that down-up DWs are initially displaced towards positive $x$-values, whereas up-down DWs remain trapped at their initial locations for $0 < t < 2$ ns, *i.e.*, when the out-of-plane field is negative ($B_e < 0$). Down-up and up-down DWs repel each other due to the DMI[12] and therefore, they do not annihilate. When the field switches to the positive polarity ($B_e > 0$, for $t > 2$ ns), down-up DWs move back to their initial locations whereas up-down DWs jump to the adjacent energy minima. As a result, the initially coded information is perturbed after $t = 2$ ns. Note that the right up domain has increased its length ($2d$) with respect to the initial state ($d$). Therefore, the proper operation of this field-driven device requires a minimum bit size equal to $2d$, where $d$ is the period of the anisotropy profile. Moreover, a complete bit shifting requires a total time given by the duration of the bipolar field pulse. Next section demonstrates that the current-driven mechanism only requires one characteristic length $d$ to store one bit, whereas the unidirectional bit shifting can be promoted by single unipolar current pulses.

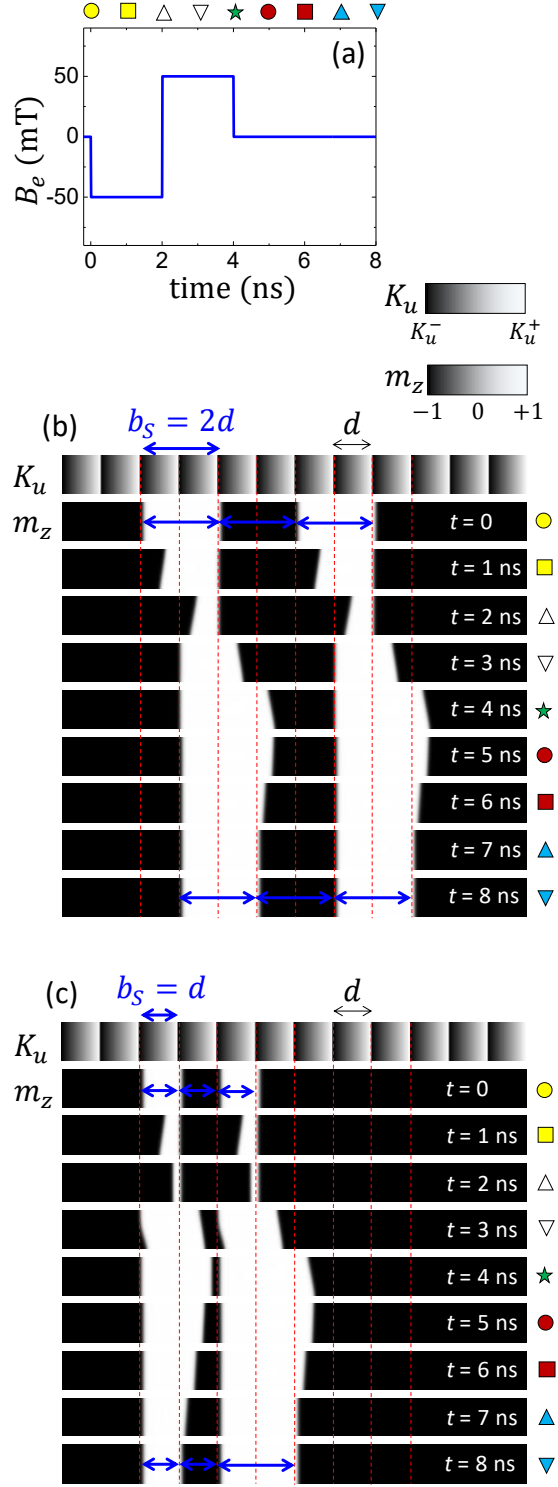

**Figure 3**. Micromagnetic results of the field-driven DW motion along a ratchet strip. (a) shows the bipolar field pulse needed to shift the bit along the positive direction of the $x$-axis. The amplitude of the bipolar field pulse is $B_e = 50$ mT, and it is negative during the first 2 ns and positive from 2 ns to 4 ns. The other two panels depict transient magnetization snapshots of the bit shifting of two bits

of size (b) $b_s = 2d$ and (c) $b_s = d$ with $d = 128$ nm. Materials parameters are those given in Sec. II. Results were obtained at zero temperature.

### III.B Current-driven case

Let us now evaluate the current-driven DW motion in the same ratchet strip as in Fig. 3. A single unipolar current pulse with amplitude $J_a = 0.8 \text{ TA/m}^2$ and duration 2 ns is applied (see Fig. 4(a)). An example of the current-driven mechanism is displayed in Fig. 4(b). As in the previous field-driven case of Fig. 3(b), three central domains (↑↓↑: 101) are separated by down-up (↓↑) and up-down (↑↓) DWs, which are initially at their equilibrium positions and at a distance equal to two characteristic lengths $d$ (bite size is $b_s = 2d$). Every DW is moved in the same $x > 0$ direction when a positive $J_a > 0$ pulse is applied, and therefore a current in the opposite direction is not required. Indeed, the DWs are displaced to the subsequent equilibrium position of the anisotropy profile. Moreover, the coherent DWs displacement can be exploited to reduce the bit size to $b_s = d$, as Fig. 4(c) shows. Contrary to the case field-driven case depicted in Fig. 3(c), now the coded bits (↑↓↑: 101) within the domains are preserved after the application of a single current pulse. Note also that even in the case of $b_s = 2d$ (Fig. 4(b)), the current-driven ratchet requires less time to promote the unidirectional bit shifting as compared to the field-driven case (Fig. 3(b)).

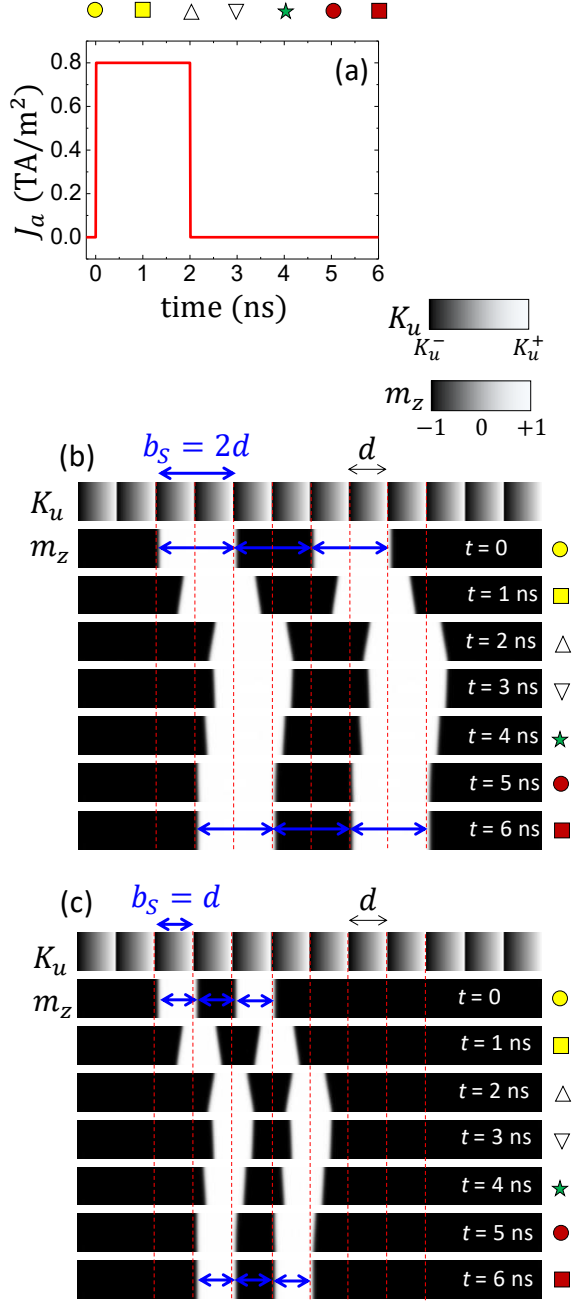

**Figure 4.** Micromagnetic results of the current-driven DW motion along a ratchet strip. (a) shows the unipolar current pulse needed to shift the bit along the positive direction of the $x$-axis. The amplitude of the pulse is $J_a = 0.8 \text{ TA/m}^2$, and its duration is 2 ns. The other two panels depict transient magnetization snapshots of the bit shifting of two bits of size (b) $b_s = 2d$ and (c) $b_s = d$ with $d = 128$ nm. Materials parameters are those given in Sec. II. Results were obtained at zero temperature.

## IV. PERFORMANCE OF CURRENT-DRIVEN DW MOTION ALONG A RATCHET STRIP

**IV.A. Micromagnetic and 1DM results. Perfect strips at zero temperature.**

The advantages of the current-driven mechanism over the field-driven case for a ratchet device have been highlighted in the previous section. Here we focus our attention to the analysis of the performance of these current-driven ratchet devices. The working principle of the current-driven ratchet device here proposed is described as follows. Each region of length $d$ can pin one DW, which in the absence of external stimuli locates approximately at any of the points where the anisotropy energy is a minimum. The region size indeed defines the bit size ($b_s = d$), and bit shifting should occur after the application of current pulses $\vec{J}(t) = J_a(t)\vec{u}_x$, since the DWs can skip the tooth of the anisotropy profile pushed forward by the driving force associated with the current. Therefore, the proper performance of the device depends on the amplitude $J_a$ of the injected current, its duration, which is referred to as the excitation time ($t_e$), and the interval between subsequent pulses, named here as the relaxing time ($t_r$). The ratio between the excitation time ($t_e$) and the total latency time ($\tau = t_e + t_r$) defines the *duty cycle* ($t_e/\tau$). Different behaviors can be observed depending on these inputs $J_a$, $t_e$ and $t_r$.

Fig. 5(a) presents the results obtained at zero temperature for three different current amplitudes $J_a$ and fixed excitation and relaxation times ($t_e = t_r = 2$ ns) corresponding to a 50% duty cycle. Micromagnetic results ($\mu M$) are depicted by dots and 1DM results with lines. A single DW located at the position of one of the anisotropy landscape minima is considered as the initial state, and two consecutive current pulses are applied. The temporal evolution of DW position is plotted when low (Fig. 5(a)), intermediate (Fig. 5(b)) and high (Fig. 5(c)) current pulses are applied. Representative micromagnetic snapshots are shown in the right panels. It can be checked that the values provided by the 1DM are in a rather good agreement with the $\mu M$ results. As it has been stated, a positive current promotes DW motion from left to right. This driving force must overcome the restoring force associated with the slopes of the anisotropy profile. If the driving force overwhelms the restoring force at each point, the DW is able to skip one or several teeth of the anisotropy profile, as long as the current is on ($t < t_e$). When the current is switched off at $t = t_e$ the DW moves backwards due to the above-mentioned restoring force, going back to the local equilibrium position at an anisotropy minimum. The current is off during a relaxing time $t_r$ which must be sufficiently long to let the DW reach such local equilibrium position. This ensures that the application of the new current pulse drives the DW in a predictable way.

Fig. 5(a) illustrates a frustrated jump attempt followed by a complete DW jump after the second current pulse ($J_a = 0.4$ TA/m$^2$). Initially, the DW is driven by the current a distance shorter than $d$. Then, the DW does not skip the anisotropy tooth but reverses its motion due to the anisotropy profile for $t_e < t < t_e + t_r$. If the relaxation time $t_r$ was sufficiently long, the DW would recover its starting position at equilibrium. However, the time $t_r = 2$ ns is in this case rather short, and the DW acquires an intermediate position on the slope when the second current pulse is applied at $t = t_e + t_r = 4$ ns. The subsequent jump is caused by the fact that the DW starts its dynamics from this non-equilibrium intermediate position after the first pulse, *i.e.*, eventual (non predictable) bit shifting occurs during the application of the second current pulse, since DWs may skip the anisotropy tooth from such intermediate positions. This unpredictable behavior is completely undesirable for applications. Conversely, high current amplitudes can make DWs eventually advance two or more bits after every single current pulse, instead of only one. Fig. 5(c), which corresponds to a current pulse with $J_a = 1.1$ TA/m$^2$, displays this behavior. The application of the first current pulse results in a double bit shift, and the second pulse also promotes a double bit shift. Finally, Fig. 5(b) depicts the case when the amplitude of the pulse is adequately high as to promote the DW displacement over a distance larger than $d$ but shorter than $2d$ ($J_a = 0.6$ TA/m$^2$). For this amplitude of the current pulse DWs do not return to the starting position after the application of each current pulse, but to the subsequent equilibrium position at a distance $d$ provided the relaxing time is sufficiently long. This situation is achieved with a high confidence in this example, and corresponds to the desired behavior, where a single bit is shifted only one tooth after every single current pulse. In other words, single DW jump occurs and the positions of the DWs always match an equilibrium state prior to the application of the next current pulse. As it can be checked in left graphs of Fig. 5, the 1DM model provides us with adequate predictions of the time evolution of DWs, in agreement with full $\mu M$ results.

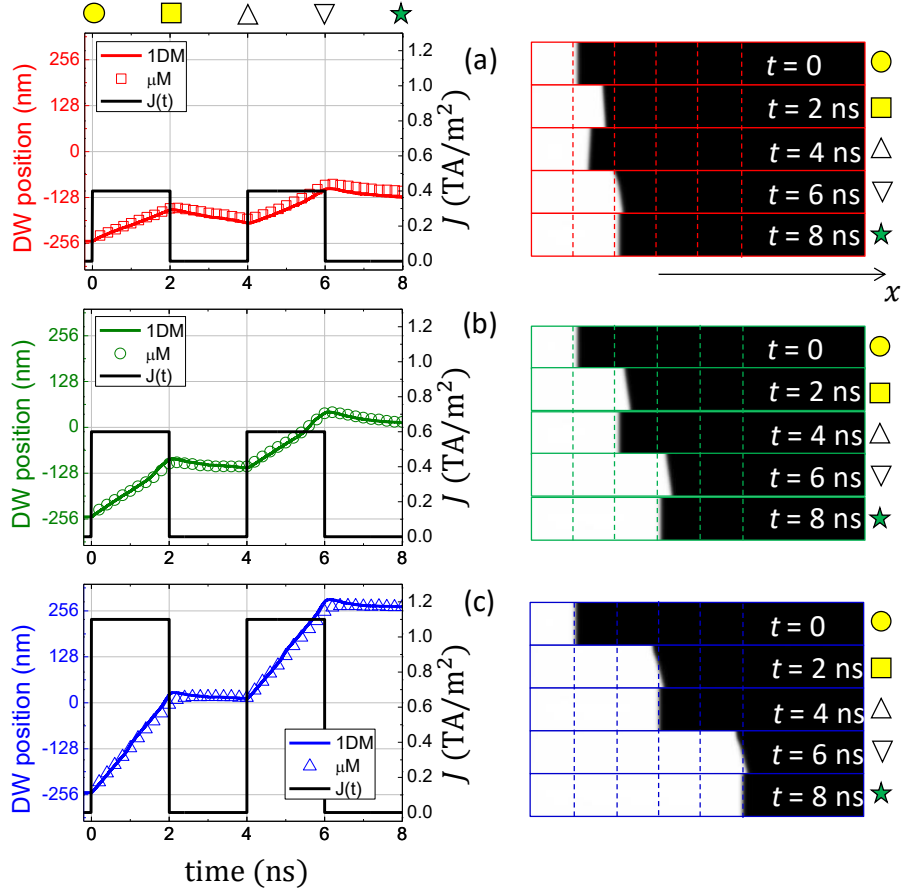

**Figure 5**. DW position as a function of time for two successive current pulses of $t_e = 2$ ns excitation time and $t_r = 2$ ns relaxing time. Three different current amplitudes at zero temperature are considered: (a) $J_a = 0.4$ TA/m$^2$, (b) $J_a = 0.6$ TA/m$^2$, and (c) $J_a = 1.1$ TA/m$^2$. In the left graphs, points correspond to full Micromagnetic results ($\mu M$) and lines are the results of the 1DM. Micromagnetic snapshots at the times when the current is switched on and off are shown in the right panels. Dashed lines in the snapshots indicate the peaks of the anisotropy profile.

### IV.B. Thermal effects: Joule heating and operation range

Former results were computed at zero temperature ($T = 0$). However, real devices should work at finite temperature and therefore, the effect of thermal fluctuations needs to be evaluated. Moreover, DWs are moved by an electric current and consequently, the effect of Joule heating may increase this temperature. Accordingly, thermal effects have been included in our results by means of the procedure described in Sec. II, whereas the temperature increase due to the Joule heating was computed by COMSOL[35] simulations of the full

HM/FM/Oxide stack (see its cross section in Fig. 6(a)). The thermal parameters of the different layers are given in Fig. 6(b), where $\sigma$ is the electrical conductivity, $C_p$ the heat capacity, $k_{th}$ is the thermal conductivity, and $\rho_m$ is the density. A thin silicon oxide layer (SiO) is assumed to be formed at the interface between the Si wafer and the Pt layer, which introduces an interfacial thermal resistance ($R_{th}$). Further details of this Joule heating evaluation are given in our previous works (see Refs. 36,37). The study indicates that within the range of current amplitudes considered along this work the temperature rises from room temperature ($T = 300$ K) following an exponential law, with a characteristic time of a few tenths of nanoseconds. After that, when the current pulse is switched off, the temperature decreases in identical fashion. As to exemplify these statements, the temporal evolution of the temperature is shown in Fig. 6(c) for three current pulses with $t_e = 2$ ns and different amplitudes $J_a$. The temporal evolution of the temperature in the FM layer obtained by COMSOL simulations can be analytically described by $T(t) = T_0 + \beta J_a^2[1 - \exp(-t/\tau_{th})]$ during the current pulse $0 < t < t_p$, and $T(t) = T_0 + \beta J_a^2 \exp(-(t - t_p)/\tau_{th})$ after it ($t \geq t_p$), where $T_0 = 300$ K, $\beta = 2.1 \times 10^{-23}$ K/(A/m$^2$), and $\tau_{th} = 0.4$ ns. Note that the maximum temperature reached at the end of the current pulse ($T_{max} = T(t = t_e)$) is well below the Curie temperature of conventional ferromagnetic materials. In particular, it is very well below the Curie temperature of the Co ($T_C \sim 1400$ K)[38]. In Fig. 6(d) we plot the maximum temperature $T_{max}$ reached at the end of the current pulse as a function of the current pulse amplitude ($J_a$). Although $T_{max}$ scales with $J_a^2$, the temperature remains well below the Curie threshold for the current amplitudes considered in the present study. Moreover, we have compared the DW dynamics along the ratchet device under a current pulse of $J_a = 0.65$ TA/m$^2$ and $t_e = 2$ ns, assuming both a constant room temperature ($T = 300$ K) and a temperature rise due to the Joule heating as described above (not shown). We checked that the dynamics is not significantly perturbed by Joule heating, and consequently, Joule heating does not constitute a drawback in the evaluated devices within the range of current amplitudes and pulse lengths considered here.

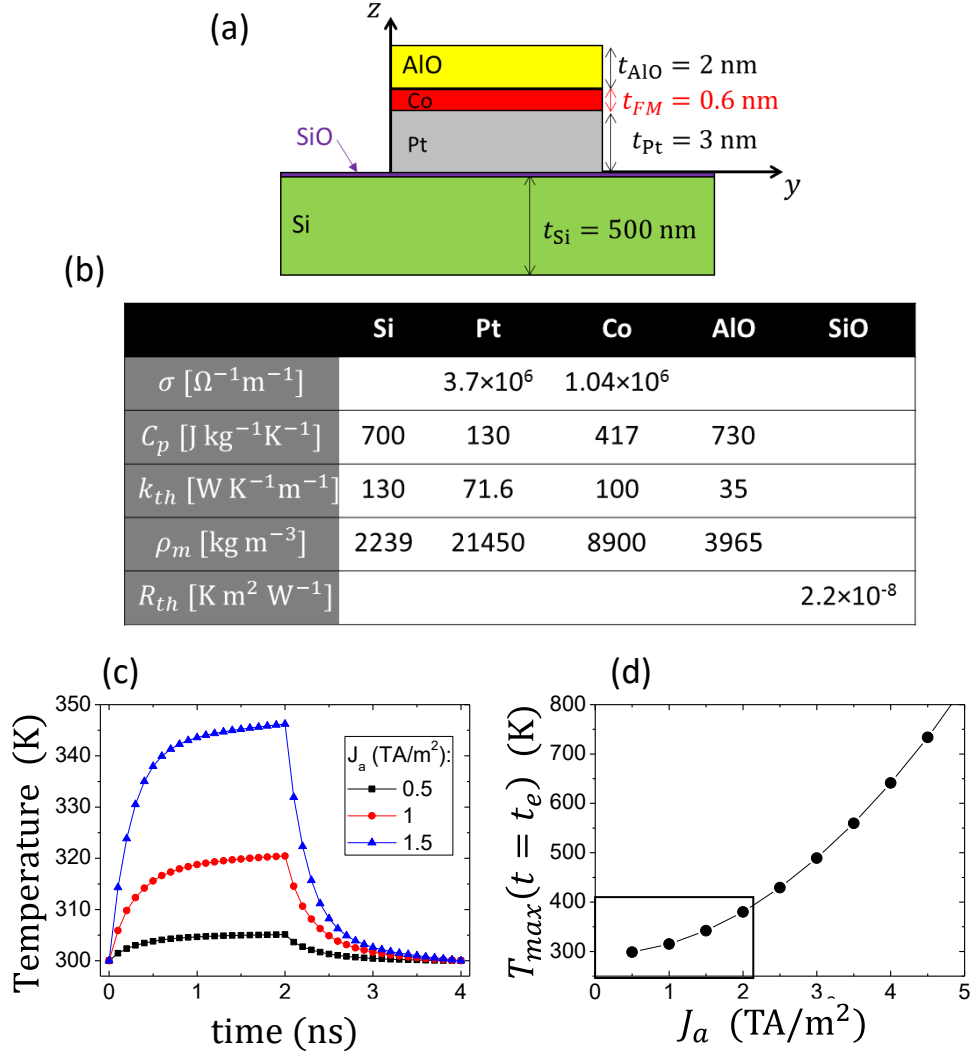

**Figure 6**. Temperature dynamics under current pulses as computed by COMSOL simulations. (a) Cross section of the multilayer studied by COMSOL simulations for the heat transport under current pulses. (b) Thermal parameters used in the simulations. (c) Temporal evolution of the temperature for three current pulses with $t_e = 2$ ns and different amplitudes ($J_a$). (d) Maximum temperature reached at the end of the current pulse ($T_{max} = T(t = t_e)$) as function of $J_a$. The box indicates the proper operation range of the considered device.

Although Joule heating seems to be negligible for the short and current pulses needed for the proper performance of these ratchet devices, thermal fluctuations can significantly reduce the reliability of the device even at fixed temperature. In order to study this reliability, the concept of proper *operating range* of the device can be introduced. This concept defines the range of values of the current amplitude $J_a$ for given times $t_e$ and $t_r$ that promote a single DW jump after the application of one current pulse. Indeed, the wider the operating range is, the more reliable the device becomes. As an example, Fig. 7(a) depicts the probability of one single DW jump after the application of one current pulse, obtained from both $\mu M$ simulations and the 1DM, as a function of the current amplitude $J_a$. Within this context, a probability of one means one single DW jump for every realization, for a total of twenty realizations at room temperature ($T = 300$ K). Three different excitation times, $t_e$: 1 ns, 1.5 ns, 2 ns, are considered for a 50% duty cycle ($t_e = t_r$). The curves neatly define for each excitation time a range of applied currents leading to single DW jumps, *i.e.*, the proper operating range. Importantly, this range increases with decreasing pulse lengths. It can be checked from the graphs in Fig. 7 the remarkable agreement between the results provided by full $\mu M$ simulations and the 1DM even at finite temperature, in particular within the area of interest which defines the proper operating range. Actually, the use of the 1DM permits us to easily extend this study over the excitation times ($t_e$) along with the current amplitudes ($J_a$). The color map in Fig. 7(b) collects this study. As in previous examples, room temperature ($T = 300$ K) is considered. As previously said, the proper operating range increases with decreasing excitation times. Besides, the shorter excitation times ($t_e$) are considered, the higher amplitudes ($J_a$) are required to achieve single DW jumps. The highest current amplitudes suitable for this device might be then limited by the appearance of DW tilting[22,28] or other effects such as a non-negligible Joule heating for currents higher than 2 TA/m$^2$. In any case, thermal fluctuations are not as intense as to endanger the performance of this device at room temperature.

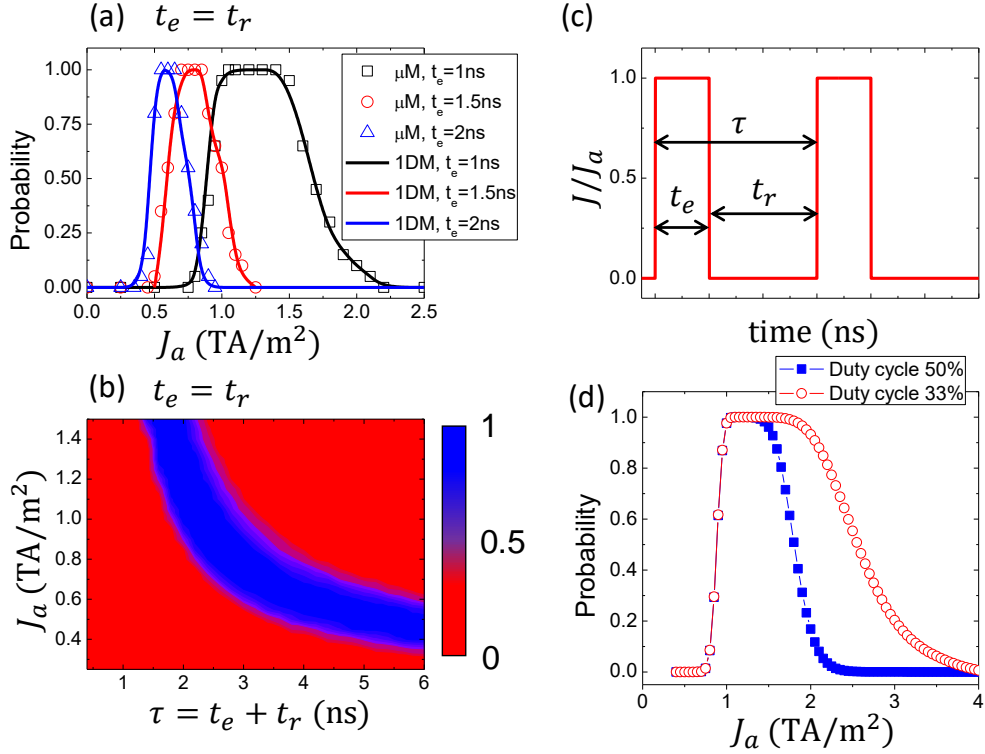

**Figure 7**. (a) Probability of single jumps of one DW after the application of one pulse. The probability is given as a function of the current amplitude for pulses with $t_e = 1$ ns, $t_e = 1.5$ ns, and $t_e = 2$ ns and are computed from full $\mu M$ simulations (symbols) and the 1DM (lines). (b) Color map computed by using the 1DM showing the probability of single DW jumps at room temperature as a function of the current amplitude $J_a$ and latency time $\tau = t_e + t_r$, the latter defined as the total time required to move one bit. A duty cycle of 50% is considered in both cases. (c) Applied current as a function of current amplitude $J_a$, pulse length $t_r$, and period $\tau = t_e + t_r$. (d) Probability of single jumps of one DW promoted by one current pulse with 50% and 33% duty cycles. Probabilities are calculated over 100000 jump attempts by means of the 1DM. The excitation time is $t_e = 1$ ns in both cases, while the relaxing times are $t_r = 1$ ns for the 50% duty cycle and $t_r = 2$ ns for the 33% duty cycle.

With the aim of extending the operating range of this device, the following strategy can be proposed. Since the DW requires a minimal relaxing time ($t_r$) to reach an equilibrium position, this time can be lengthened, then reducing the duty cycle. In that way, a 50% duty cycle means that $t_e = t_r$ while a 33% duty cycle stands for $2t_e = t_r$. In order to provide a detailed statistics of the reliability of the proposed device, Fig. 7(d) compares the probability at finite temperature ($T = 300$ K) of single DW jump after one current pulse over 100000 attempts (see the previous definition of probability equal to 1) calculated by means of the 1DM. The two mentioned duty cycles, 50% and 33%, are considered with fixed excitation time $t_e = 1$ ns. The results for current pulses with 50% duty cycle reveal a shorter range of currents resulting in proper working, as making the device more vulnerable to thermal fluctuations. Nevertheless, a 33% duty cycle ensures proper working for a wider range of applied currents, which increases the reliability of the device.

### IV.C. The effect of disorder

Results discussed in previous sections were evaluated assuming perfect samples, without defects nor imperfections. Here we show that current-driven DW motion along a ratchet (unidirectional) device is also reliable even under realistic conditions. The DW motion under current pulses for perfect strips (with neither edge roughness nor bulk defects) is compared with a more realistic description. Such realistic conditions were evaluated by assuming that the strip has some edge roughness and bulk defects in the form of grains (see Sec. II.A). In Fig. 8, the perfect and the realistic cases are compared under two consecutive current-pulses with $t_e = 2$ ns and $t_r = 4$ ns (which correspond to a duty cycle of 33%). Room temperature ($T = 300$ K) and current pulse amplitudes of $J_a = 0.65$ TA/m² were considered in both cases. Note that both thermal fluctuations and disorder make the DW profile rough as compared to the perfect case (no disorder) at zero temperature (see Fig. 5, for example). Anyway, the results presented in Fig. 8 indicate that realistic conditions (with edge roughness and bulk defects in the form of grains) do not significantly deviate from the ideal (perfect) case, and therefore, the device working principle is not compromised by realistic conditions.

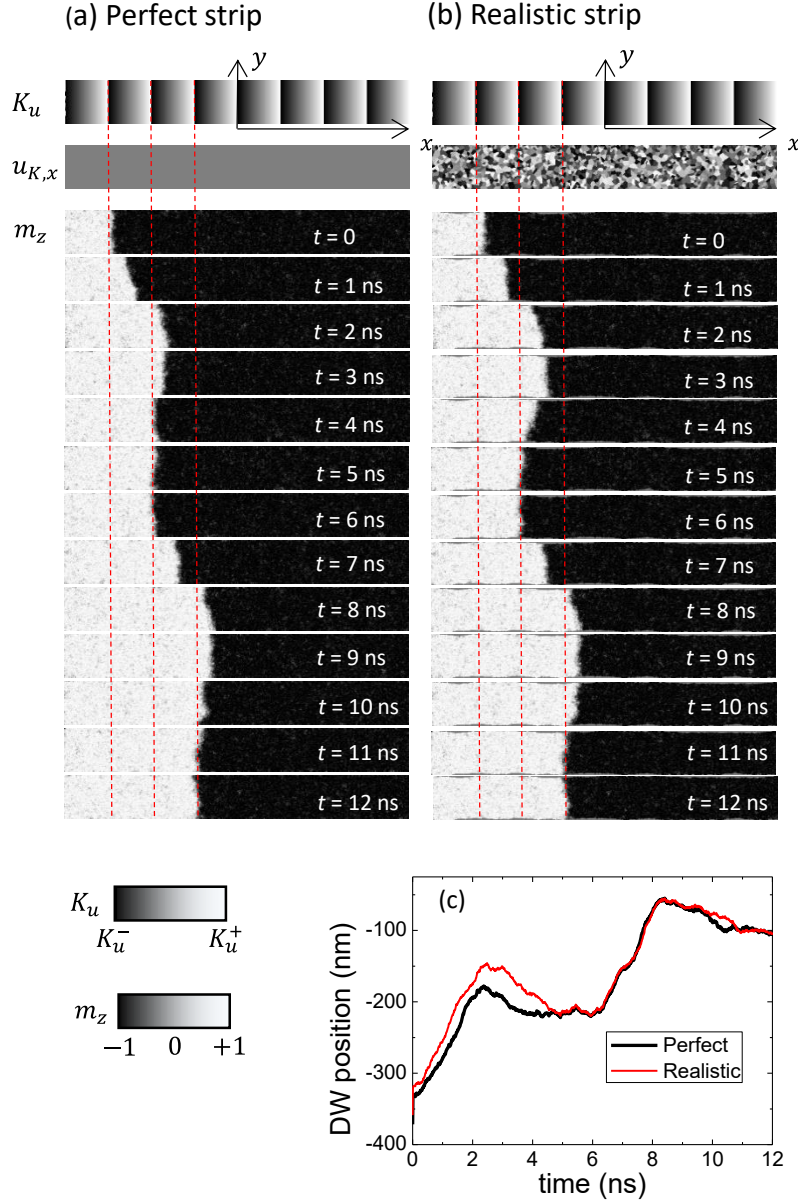

**Figure 8**. Micromagnetic analysis of the current-driven DW motion along a ratchet strip: (a) perfect *vs* (b) realistic strips. $\mu M$ simulations were performed with the inputs given in Sec. II and at room temperature. Here the amplitude of the current pulse is $J_a = 0.65 \text{ TA/m}^2$ and two consecutive current-pulses of $t_e = 2$ ns and $t_r = 4$ ns are applied. The anisotropy profile and the $x$-component of the uniaxial easy axis are shown on top of the snapshots. In (b), bulk grains and edge roughness are taken into account as described in Sec. II.A. Snapshots from top to bottom correspond to different magnetization states at consecutive instants of time. The temporal evolution of the DW position is plot in (c).

**IV.D. Material parameters and variable DMI**

The performance of the proposed current-driven ratchet device can be dependent in general on the material parameters. For example, studies (not shown here) on the dependence of this performance on the Gilbert damping ($\alpha: 0.1, ..., 0.5$) or other extreme values of the PMA constant ($K_u^-, K_u^+$) have also been carried out, not leading to severe qualitative modifications of this performance. Moreover, these parameters can be adequately tuned in most cases to fix the most convenient set of them for the desired characteristics. In that way, the 1DM becomes a useful tool to determine the proper parameters.

The influence of the DMI parameter deserves particular attention. In fact, a constant DMI parameter has been considered along this work. Although it has been recently demonstrated that PMA and DMI can be tuned almost independently by adjusting irradiation energies and doses[19], a small variation of the DMI may be expected when tailoring the PMA by ionic irradiation. Here, a linear variation of the DMI parameter ($D(x)$), analogous to that of the PMA ($K_u(x)$), has been also considered. In this way, the DMI parameter changes linearly from $D^- = 0.8$ mJ/m² to $D^+ = 1.0$ mJ/m² in a periodic fashion with period $d = 128$ nm. Note that within this range, the DMI parameters are still sufficiently high as to give rise to the formation of chiral Néel DWs, so that efficient current-driven DW movement by means of the SHE is still guaranteed. $\mu M$ simulations have been carried out to compare the results obtained both with constant DMI basis and with the considered linear variation of the DMI (see Sec. II.A for details). 33% duty cycle current pulses are applied with an amplitude $J_a = 1.2$ TA/m², and the probability of single DW jump after each current pulse is analyzed in both cases at finite temperature. The results are plotted in Fig. 9, which indicates that no drastic changes are found in the case of variable DMI with respect to the case of constant DMI, and therefore the working principle is still preserved.

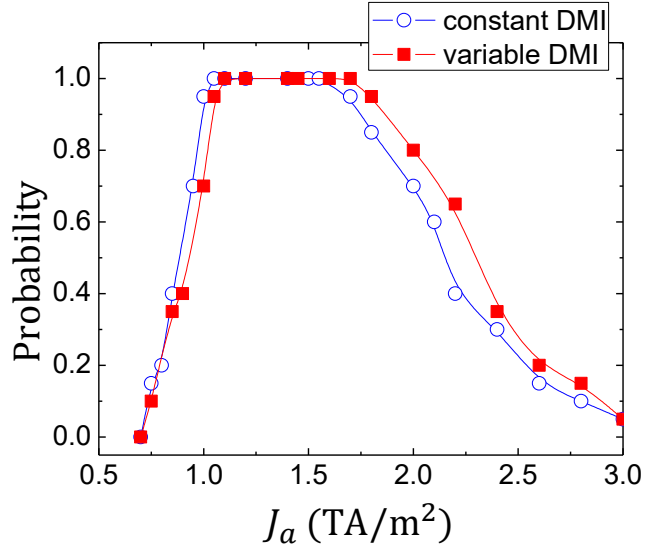

**Figure 9**. Micromagnetically computed probability of single jumps of one DW over twenty attempts with constant DMI parameter (open symbols) and with a linear variation of the DMI parameter (filled symbols). In the first case the DMI parameter is fixed ($D = 1.0$ mJ/m$^2$), whereas in the second one $D = D(x)$ changes linearly from $D^- = 0.8$ mJ/m$^2$ to $D^+ = 1.0$ mJ/m$^2$ in a periodic fashion with the same period of the PMA constant ($d = 128$ nm).

## V. CURRENT-DRIVEN DW MOTION ALONG A BI-DIRECTIONAL DEVICES

The ratchet PMA profile (Fig. 2(b)) was designed to avoid DW backward movement in a field-driven basis. However, this results in a unidirectional motion of trains of DWs, making this device less flexible as compared to other racetracks memories devices intended for bi-directional DW motion[1,2]. Our methods ($\mu M$ and 1DM) can be straightforwardly adapted for a perpendicular profile ($K_u = K_u(x),$) designed now to promote the bi-directional shifting of trains of DWs. This can be easily performed by considering a periodic triangular anisotropy profile, which consists on parts where $K_u(x)$ increases linearly up to $K_u^+$ over a distance $d_1 = d/2$, and from there, $K_u$ decreases down to $K_u^-$ with the same slope within an identical distance ($d_2 = d/2$). The analytical expression for this PMA profile $K_u = K_u(x)$ can be obtained from Eq. (5) with $d_1 = d_2 = d/2$, and it is shown in top graph of Fig. 10(a). The corresponding effective field due to the PMA profile ($r(q) = H_{PMA}(q)/H_{PMA}^0$) is obtained from Eq. (8) and it is represented in the bottom graph of Fig. 10(a). We have considered a single DW initially located at one of these energy minima, *i.e.*, at a point where $K_u(q) \approx K_u^- (q(t = 0) \approx -d)$, and positive current pulses are applied with fixed duration $t_e = 2$ ns and two different amplitudes: $J_a = 0.5$ TA/m$^2$ and $J_a = 0.6$ TA/m$^2$ (see Fig. 10(b)). The temporal evolution of the DW position is depicted in Fig. 10(c), where open symbols corresponds to $\mu M$ results and solid lines were obtained from the 1DM. For $J_a = 0.5$ TA/m$^2$, the DW does not overcome the energy barrier at the end of the pulse, and it relaxes back to its initial location for $t > t_e = 2$ ns. On the contrary, a current pulse with $J_a = 0.6$ TA/m$^2$ is enough to overcome the energy barrier (located at $q = -d + d/2$) during the current pulse ($0 < t_e < 2$ ns). Therefore, this DW relaxes to the adjacent energy minima (at $q = -d + d = 0$), where it rests for $t > 6$ ns. As it is clearly seen, both $\mu M$ and 1DM results are again in good quantitative agreement.

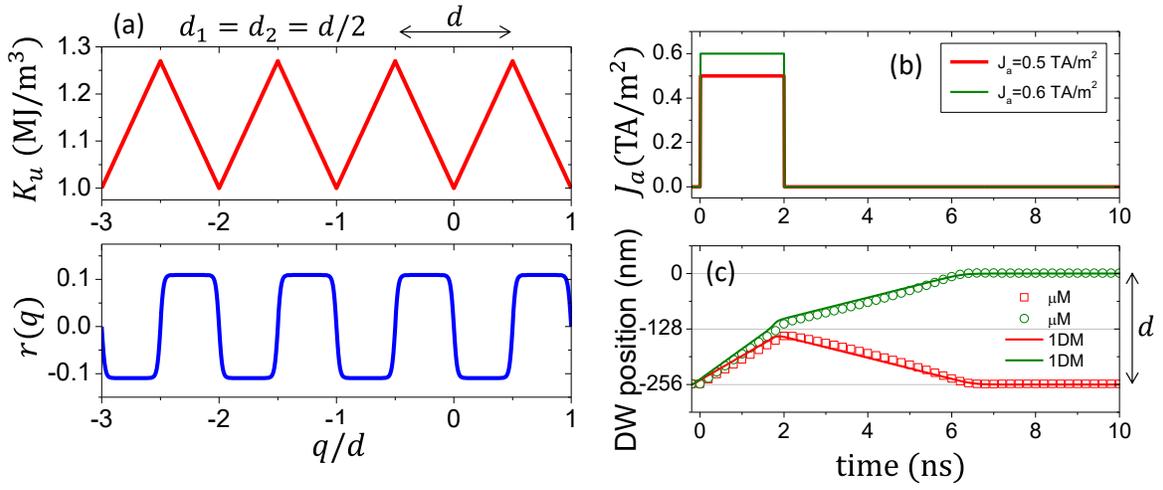

**Figure 10.** (a) Periodic triangular PMA profile ($K_u(x)$, Eq. (5)) to promote DW motion in bi-directional devices. Here $K_u^- = 1.0$ MJ/m$^3$, $K_u^+ = 1.27$ MJ/m$^3$, $d_1 = d_2 = d/2$, with $d = 256$ nm. The bottom graph represents the corresponding $r(q) = H_{PMA}(q)/H_{PMA}^0$ along the strip axis (Eq. (8)). (b) Two current pulses with fixed duration $t_e = 2$ ns and two different amplitudes: $J_a = 0.5$ TA/m$^2$ (red) and $J_a = 0.6$ TA/m$^2$ (green). (c) Temporal evolution of the DW position initially located at under the current pulses. Open symbols correspond to $\mu M$ results and solid lines were obtained by solved the 1DM Eqs. Perfect samples (without imperfections) and zero temperature conditions were assumed in both models.

In this bi-directional device, an isolated bit of information is coded within a domain magnetized up (↑, white color in magnetic snapshots of Fig. 11) bounded by two adjacent DWs placed at two energy minima. *i.e.*, at points where $K_u(x) \approx K_u^-$. This domain then includes a position where a maximum of the anisotropy locates equally distant to the domain edges where down-up (↓↑) and up-down (↑↓) DWs are. The bit size is therefore $b_S = d =$ $2d_1$. Contrary to the unidirectional ratchet device, this PMA profile (Fig. 10(a)) allows the bi-directional current-driven DW motion. In order to show it, the current-driven motion of two DWs under two consecutive current pulses of opposite polarities was evaluated. Positive current pulses ($J_a > 0$) drive both down-up (↓↑) and up-down (↑↓) DWs along the positive direction ($x > 0$), whereas negative current pulses ($J_a < 0$) drive both down-up and up-down DWs along the negative direction ($x < 0$). One example of this bi-directional functionality is shown in Fig. 11, which was obtained by means of full $\mu M$ simulations. The current amplitude is $J_a = 0.8 \text{ TA/m}^2$, and the pulse duration is $t_e = 2 \text{ ns}$. Note that such a bi-directional functionality cannot be achieved in the field-driven case, due to the simple fact that adjacent DWs move along opposite directions under a given polarity of the field. Our modeling suggests these systems as promising platforms to develop DW-based devices which can be efficiently driven back and forth by short current pulses, and where the DW position can be controlled by the periodic modulation of the PMA.

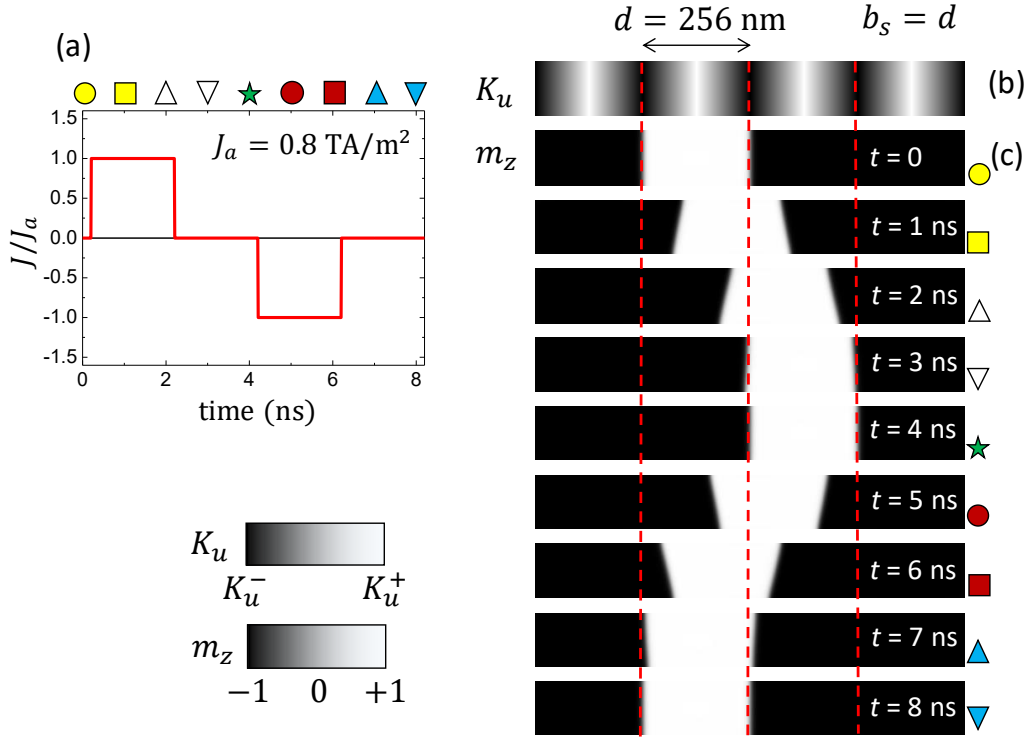

**Figure 11**. Micromagnetic analysis of the current-driven DW motion in bi-directional devices. A perfect strip at zero temperature is considered. The perpendicular anisotropy periodically increases and decreases linearly between $K_u^- = 1.1 \text{ MJ/m}^3$ and $K_u^- = 1.27 \text{ MJ/m}^3$ within a distance $d_1 = d_2 = d/2 = 128 \text{ nm}$. The rest of inputs are those given in II.A. The bit size, defined as the distance

between two adjacent energy minima, is $b_s = d$. The anisotropy profile along the ferromagnetic strip axis is shown in (b), while the two current pulses ($J_a = 0.8 \text{ TA/m}^2$ and $t_e = 2$ ns) injected to promote forward and backward longitudinal displacement of the two DWs are plotted in (a). Transient snapshots of consecutive magnetic states are shown in (c).

## VI. CONCLUSIONS

The DW dynamics along ferromagnetic strip with periodically modulated perpendicular magnetic anisotropy has been analyzed. The work presents a theoretical study of the current-driven ratchet memory as a DW-based magnetic memory. This study has been carried out by means of full $\mu M$ simulations along with the 1DM, showing that both approaches are in good agreement. The 1DM was developed for a general periodic PMA profile: a new term ($H_{PMA}(q)$) in the 1DM dynamics equations. Such a local PMA field accounts for the space dependent anisotropy profile, and forces the DW to travel the closest accessible PMA minimum in the absence of driving force. The work shows how both the bit density and shifting speed are notably enhanced if the memory device works on the current-driven basis as compared with the field-driven scheme. This enhancement roots in three main differences between the two driving forces. First of all, single pulses with a given polarity (unipolar) are required in the case of the current-driven ratchet, whereas bipolar pulses are needed in the field-driven case. Secondly, the duration of the current pulses to achieve a single DW jump over a single anisotropy slope (tooth) is significantly shorter than the duration of the needed bipolar field pulses. Moreover, the bit size ($b_s$), determined by the minimum distance between two adjacent DWs, is smaller in the current-driven case than in the field driven case. Note that each bit occupies two teeth in the field-driven ratchet ($b_S = 2d$), whereas a bit only needs a single tooth in the current-driven case ($b_S = d$). Consequently, the current-driven architecture allows higher density packed information in these devices. Additionally, its performance may even be improved by a fine tuning of the pulse and relaxing times. In this way, the best results are obtained when the combined effect of the times of both the applied current pulse and the relaxing interval leads every DW from an equilibrium state to another one at the adjacent tooth of the anisotropy landscape, the former time by promoting only one single jump and the latter by allowing the DW to recover this new equilibrium position before the next pulse is applied. Finally, our study opens a new promising door to other devices with tuned anisotropy. In particular, not only ratchet-like but also current-driven bi-directional devices can be developed.

**ACKNOWLEDGMENTS**

This work was supported by project WALL, FP7- PEOPLE-2013-ITN 608031 from the European Commission, projects MAT2014- 52477-C5-4-P, MAT2017-87072-C4-1-P, and MAT2017-90771-REDT) from the Spanish government, and projects SA282U14 and SA090U16 from the Junta de Castilla y Leon.